# Energy Efficient and High Performance Current-Mode Neural Network Circuit using Memristors and Digitally Assisted Analog CMOS Neurons

Aranya Goswamy[1], Sagar Kumashi[1], Vikash Sehwag[1], Siddharth Kumar Singh[1], Manny Jain[1], Kaushik Roy[2], Mrigank Sharad[1]

[1]Department of Electronics and ECE, IIT Kharagpur, West Bengal, India
[2]School of Electrical Engineering, Purdue University, West Lafayette, IN, USA

**Abstract: Emerging nano-scale programmable Resistive-RAM (RRAM) has been identified as a promising technology for implementing brain-inspired computing hardware. Several neural network architectures, that essentially involve computation of scalar-products between input data vectors and stored network 'weights' can be efficiently implemented using high density cross-bar arrays of RRAM, integrated with CMOS. In such a design, the CMOS interface may be responsible for providing input excitations and for processing the RRAM's output. In order to achieve high energy efficiency along with high integration density in RRAM based neuromorphic hardware, the design of RRAM-CMOS interface can therefore play a major role. In this work we propose design of high performance, current mode CMOS interface for RRAM based neural network design. The use of current mode excitation for input interface and design of digitally assisted current-mode CMOS neuron circuit for the output interface is presented. The proposed technique achieve ~10x energy as well as performance improvement over conventional approaches employed in literature. Network level simulations show that the proposed scheme can achieve 2 orders of magnitude lower energy dissipation as compared to a digital ASIC implementation of a feed-forward neural network.**

## 1. Introduction

As demand on high performance computation increases, the traditional Von Neumann computer architecture becomes less efficient. In recent years, neuromorphic hardware systems have gained great attention. Such systems can potentially provide the capabilities of biological perception and information processing within a compact and energy-efficient platform. Many research activities have been carried out on neural network algorithm enhancement and/or system implementations built upon the conventional CPU, GPU, or FPGA **[1].** In recent years several device solutions have been proposed for fabricating nano-scale programmable resistive elements, generally categorized under the term 'memristor' [1-9]. Of special interest are those which are amenable to integration with state of the art CMOS technology, like memristors based on Ag-Si filaments [6-8]. Such devices can be integrated into metallic crossbars to obtain high density resistive crossbar memory (RCM) [1-8]. Continuous range of resistance values obtainable in these devices can facilitate the design of multi-level, non-volatile memory [1-3]. The Resistive-Crossbar Memory (RCM) technology has led to interesting possibilities of combining memory with computation [1-5]. RCM can be highly suitable for a class of non-Boolean computing applications that involve pattern-matching [5, 11]. Such applications employ highly memory intensive computing that may require correlation of a multidimensional input data with a large number of stored patterns or templates, in order to find the best match [11]. Use of conventional digital processing techniques for such tasks incurs high energy and real-estate cost, due to the sheer number of computations involved. Structurally, RCM can be a much closer fit for this class of associative computation. Owing to the direct use of nano-scale memory array for associative computing, it can provide very high degree of parallelism, apart from eliminating the overhead due to memory read**[2].** Associative computing of practical complexity with RCM is essentially analog in nature, as it involves evaluating the degree of correlation between inputs and the stored data. In this project we investigate the construction of a neuron circuit that can take the dot product produced by the crossbar array as its input and accordingly produce a voltage level that can be compared with a reference voltage level to produce a digital output, using a comparator. The neuron design is essential in several respect, since it needs to act as a transimpedance amplifier with low input impedance and high tolerance to variability due to process variations or mismatch.

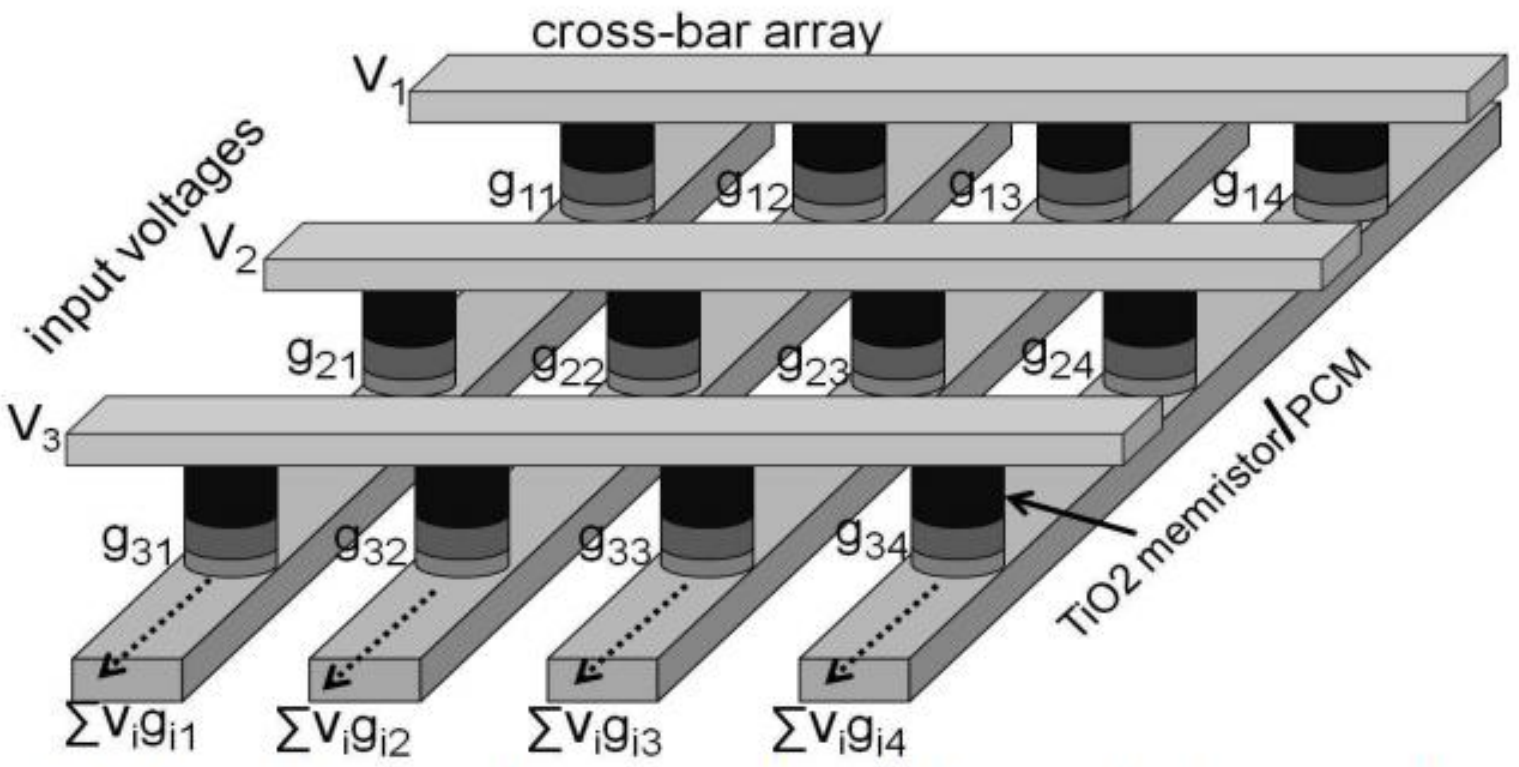

Fig. 1 A Resistive crossbar network used for evaluating correlation between inputs and stored data.

# 2. Description of Elements of Circuit

## 2.1 Regulated Cascode Transimpedance Amplifier

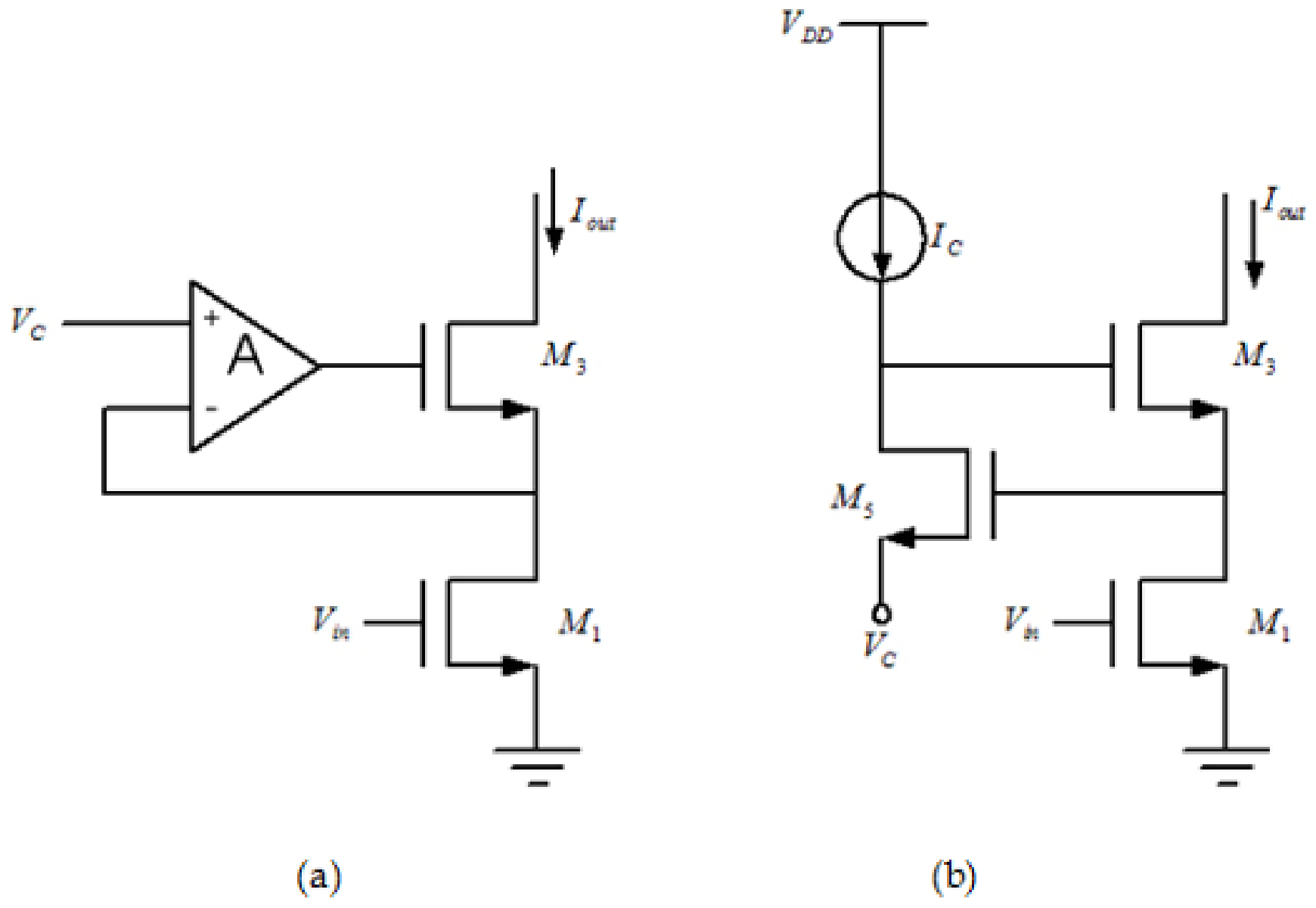

FIGURE 2.

a)Basic triode transconductor structure (b) Simple RGC triode transconductor

In Figure 2(a) regulating amplifier keeps VDS of M1 at a constant value determined by VC . It is less than the overdrive voltage of M1. The voltage can be controlled from VC so as to place M3 in current-voltage feedback, thereby increasing output impedance. The concept is to drive the gate of M3 by an amplifier that forces VDS1 to be equal to VC . Therefore, the voltage variations at the drain of M3affect VDS1 to a lesser extent because amplifiers “regulate” this voltage. With the smaller variations atVDS1 the current through M1 and hence output current remains more constant, yielding a higher output impedance [Razavi, 2001]

$$R_{out} \approx A g_{m3} r_{O3} r_{O1} \quad (9)$$

It is one of solutions using regulated cascode to replace the auxiliary amplifier in order to overcome restrictions on Figure 1. The circuit in Figure 2(b) proposed in [Mahattanakul & Toumazou, 1998] uses a single transistor, M5, to replace the amplifier in Figure 2(a). This circuit called regulated cascode which is abbreviated to RGC. The RGC uses M5 to achieve the gain boosting by increasing the output impedance without adding more cascode devices. VDS1 is calculated by follows: Assuming M5is in saturation region in Figure 2(b). It can be shown that

IC=12β5(VGS5−VT)2VGS5=VDS1−VC=2ICβ5−−−√+VT5VDS1=VC+2ICβ5−−−√+VT5 (10)

From (6)

Gm=β1VDS1=β1(VC+2ICβ5−−−−√+VT5)

. Thus, Gm can be tuned by using a controllable voltage source VC or current source IC . However, it is preferable in practice to use a controllable voltage source VC for lowering power consumption since VDS1 only varies as a square root function ofIC .

Simple RGC transconductor using a single transistor to achieve gain boosting can reduce area and power wasted by the auxiliary amplifiers. **[3]**

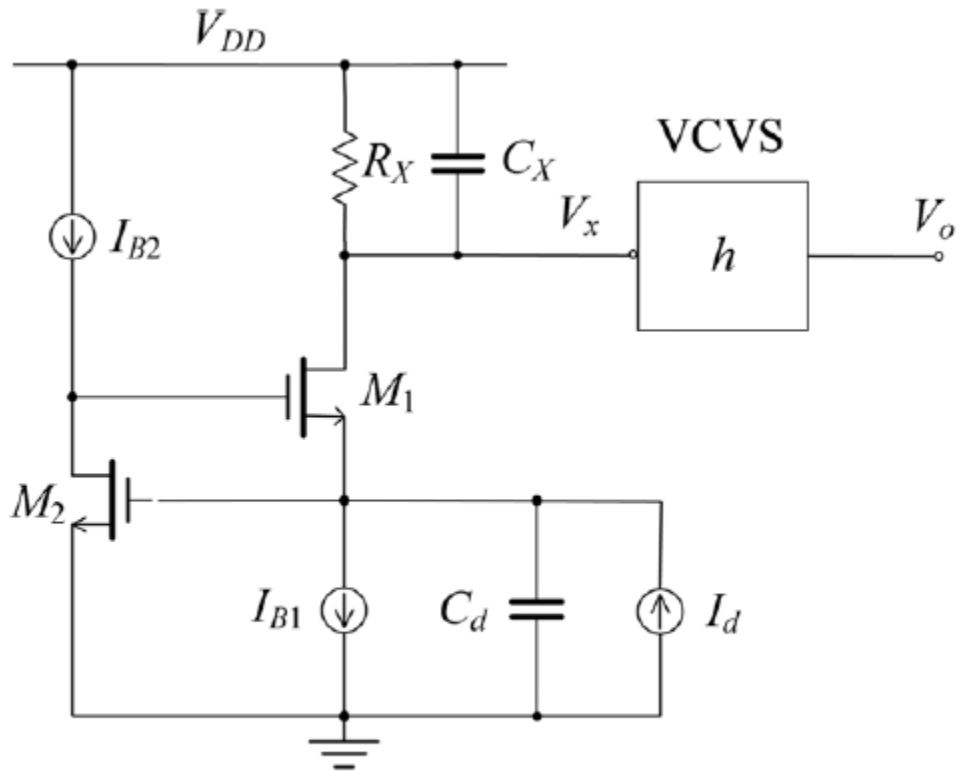

Fig. 5. Regulated cascode TIA.

The circuit in Fig. 5 [14] is usually referred to in the literature as a “regulated cascode” stage, and for this reason that is the designation that is used in this paper. However, we point out that the circuit in Fig. 5 is not derived from a cascode stage (i.e. a common-source followed by a common-gate stage); instead, it may be viewed as a common-gate stage to which a loop is added containing a voltage amplifier, which has the effect of dividing the input impedance by the amplifier gain. Furthermore, the circuit in Fig. 5 is different from a well-known high gain amplifier that is derived from a cascode stage and is properly called “regulated cascode”.

The common-source transistor M2 with active load IB2 is an amplifier stage with voltage gain A

A=gm2(r02||R0B2) (19)

where R0B2 is the incremental resistance of the load current source IB2.

The input impedance is

$$Z \approx 1/(A*gm1) \quad (20)$$

## 2.2 Successive Approximation Register

A successive approximation ADC is a type of analog-to-digital converter that converts a continuous analog waveform into a discrete digital representation via a binary searchthrough all possible quantization levels before finally converging upon a digital output for each conversion. **[3]**

Although there are many variations for implementing a SAR ADC, the basic architecture is quite simple (see Figure 1). The analog input voltage (VIN) is held on a track/hold. To implement the binary search algorithm, the N-bit register is first set to midscale (that is, 100... .00, where the MSB is set to 1). This forces the DAC output (VDAC) to be VREF/2, where VREF is the reference voltage provided to the ADC. A comparison is then performed to determine if VIN is less than, or greater than, VDAC. If VIN is greater than VDAC, the comparator output is a logic high, or 1, and the MSB of the N-bit register remains at 1. Conversely, if VIN is less than VDAC, the comparator output is a logic low and the MSB of the register is cleared to logic 0. The SAR control logic then moves to the next bit down, forces that bit high, and does another comparison. The sequence continues all the way down to the LSB. Once this is done, the conversion is complete and the N-bit digital word is available in the register.

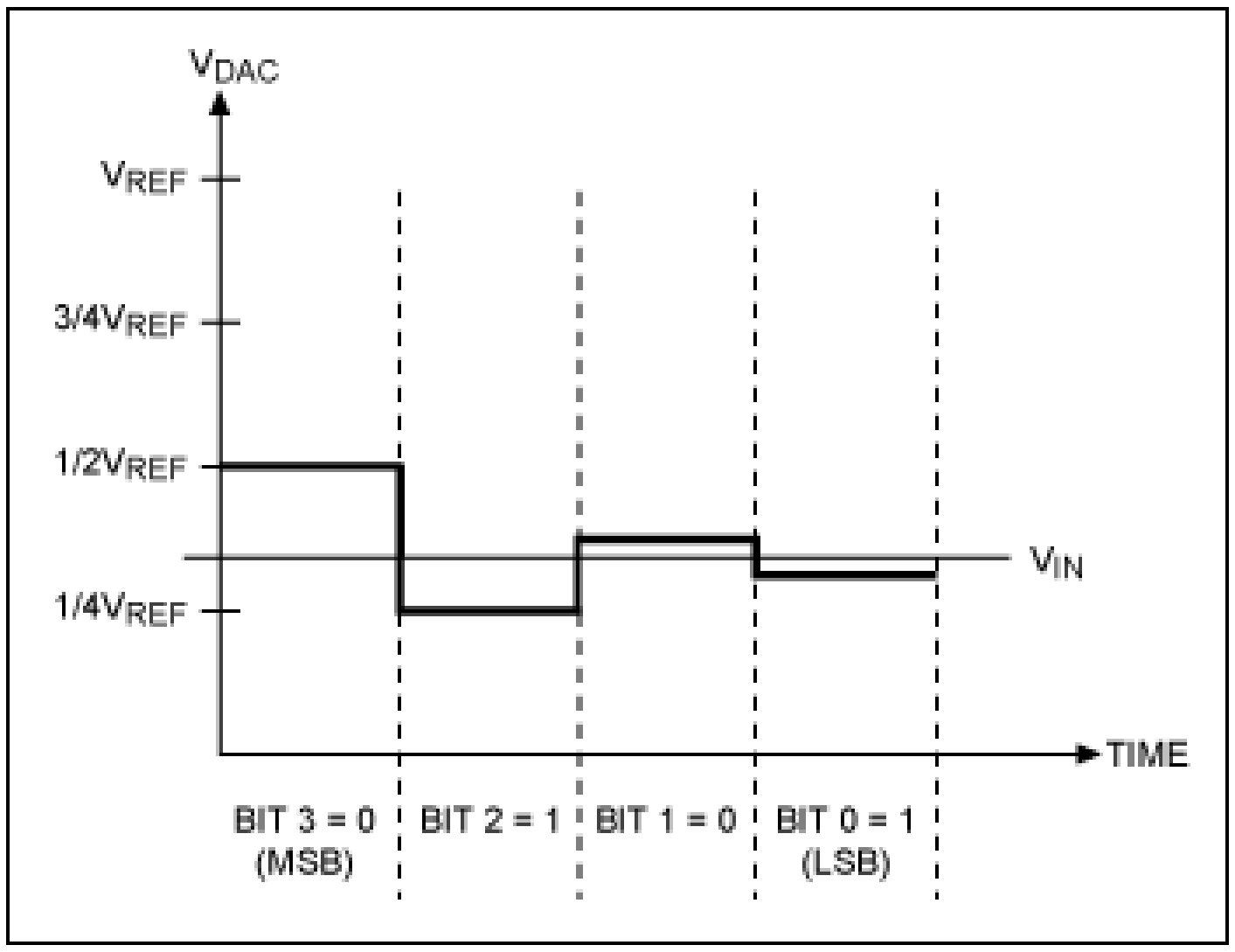

*Figure 2. SAR operation (4-bit ADC example).*

Notice that four comparison periods are required for a 4-bit ADC. Generally speaking, an N-bit SAR ADC will require N comparison periods and will not be ready for the next conversion until the current one is complete.

Mathematically, let Vin = xVref, so x in [-1, 1] is the normalized input voltage. The objective is to approximately digitize x to an accuracy of 1/2n. The algorithm proceeds as follows:

Initial approximation x0 = 0.

ith approximation xi = xi-1 - s(xi-1 - x)/2i.

where, s(x) is the signum-function(sgn(x)) (+1 for x ≥ 0, -1 for x < 0). It follows using mathematical induction that |xn - x| ≤ 1/2n. **[3]**

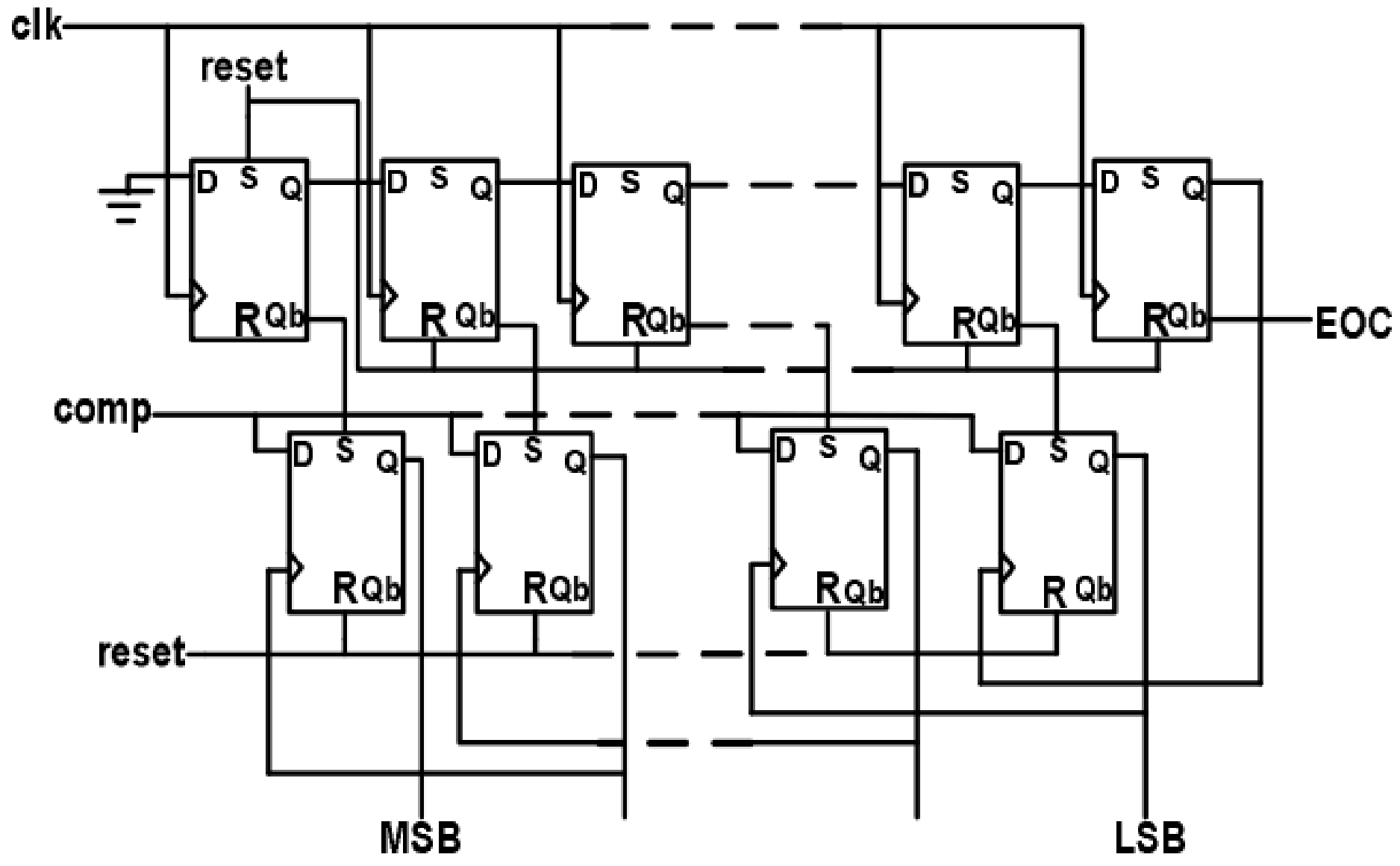

The schematic of the SAR logic consists of shift register and code shift register using D-flip flop as shown in figure ix. Initially the reset line goes low. This line controls set line of FF1 and reset lines of all other sequencer flip flops. The same reset signal also controls the reset line of code register flip flops. Q and Qb of FF1 are set to 1 and 0 respectively. Qb also controls the set line of CF1. Hence the CF1 output is forced to 1. This is the MSB bit and the weight for VFSR/2. It should be noted that since sequence register is reset initially, the set input of all the code registers flip flops except CF1 is logic 1. Hence all the other code register output states are logic0 0. We get a sequence MSB=1 and all other set to 0. The analog equivalent of this weight will be generated by the DAC. When reset goes high and clock is triggered, Q becomes 0 and FF2 outs logic high. This low to high transition of FF2 triggers or clocks the code register flip flop CF1 to store control bus value to its output. When clock runs further, the code register flip flop retains the set value as FF2 output goes to zero. This process is repeated for each of the flip flops until after N clock cycles a high state comes out of sequencer flip flop controlling the code register LSB flip flop **[4].**

## 2.3 Design of Circuit

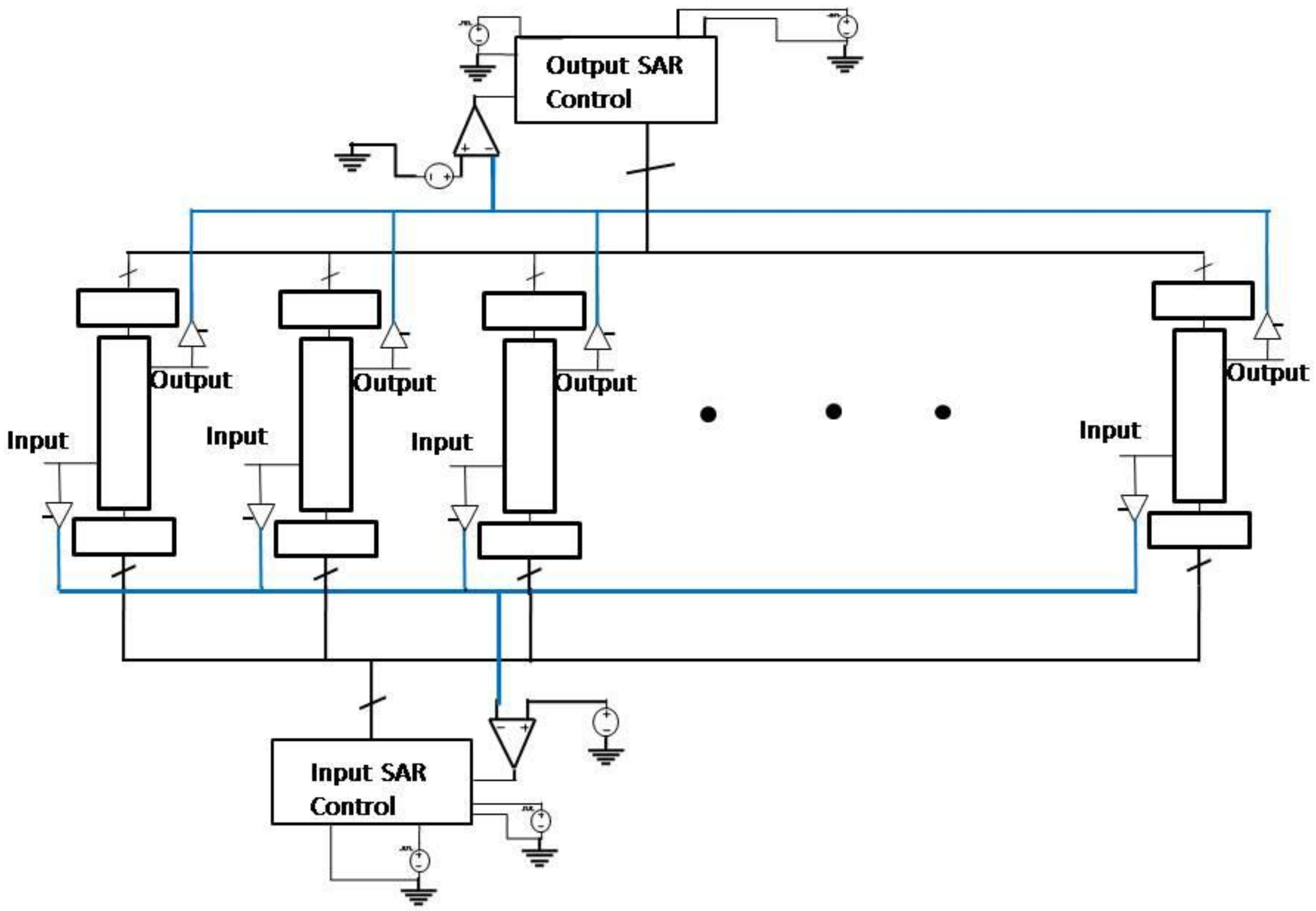

The complete circuit shows an array of neuron, each having an input and output terminal. The input terminals form the outputs of the crossbar array, thus the input is the scalar dot product of the input and the weights in the memristor array.

The SAR logic is shared both at the input and the output for the neurons. The buffers are connected to a control logic so that at any time instant only one neuron block is active. The SAR logic takes a finite time to adjust the input DC voltage of the neuron by turning on and off a particular combination of the transistors which mirror the bias current in the feedback branch. After the stabilization of one neuron input is complete, the next neuron is selected via the control logic, and the process repeats. Thus the stabilization of the whole circuit depends on the number of neurons in the array and the frequency of operation of the SAR logic block. The same technique is used to stabilize the DC point of the output.

The circuit of the individual neuron is shown below.

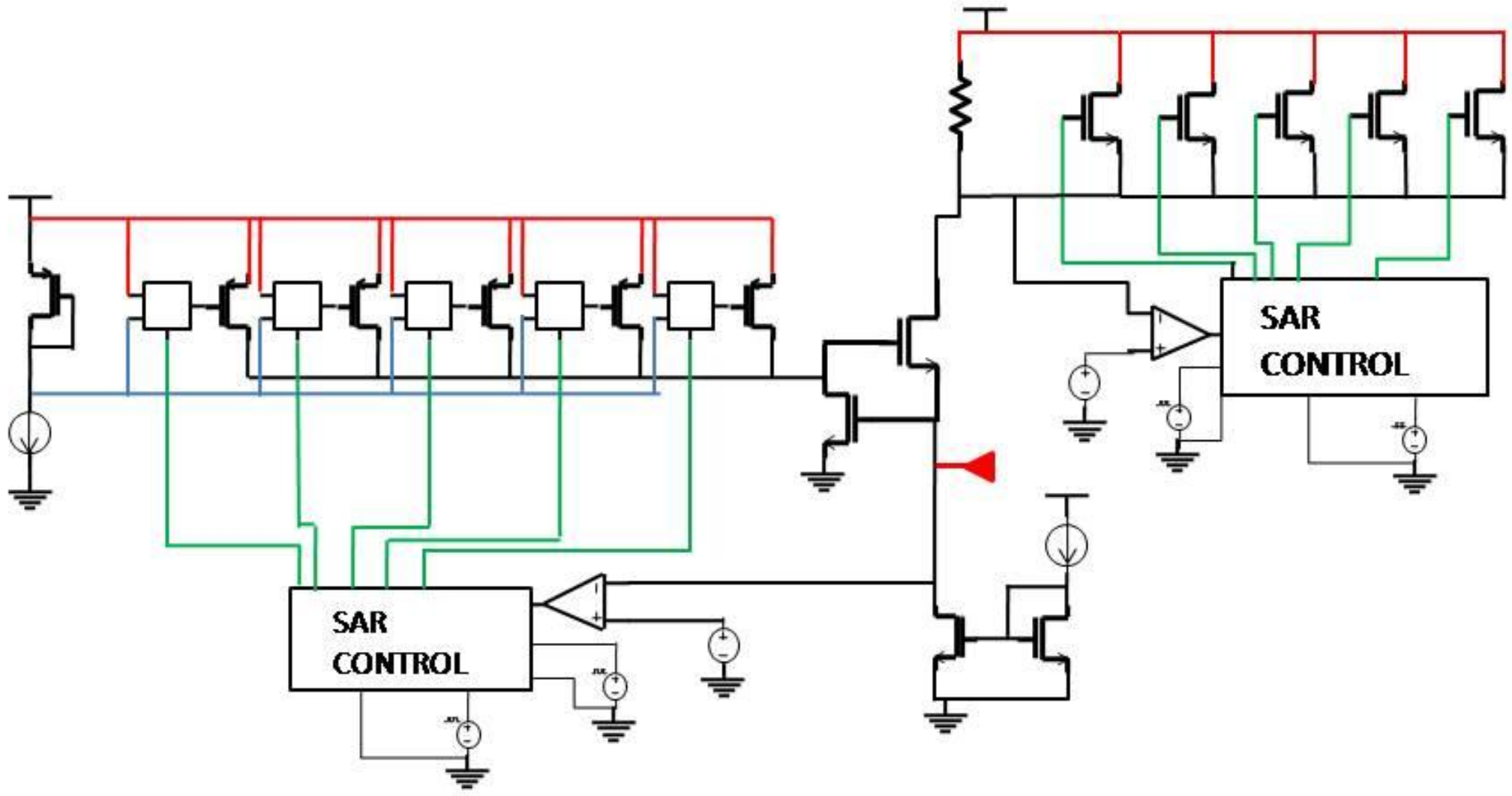

In this neuron circuit, a regulated cascade transimpedance amplifier is used, which is essentially a common gate amplifier with feedback. This reduces the input impedance and decreases the variability at the input point and also increases the output impedance. The bias current for the feedback amplifier is controlled by the binary weighted array of PMOS transistors. The SAR logic block compares the input voltage with the reference voltage and either turns on or turns off successive transistors in the array. Analog MUX is used at the gates of the PMOS transistors for proper biasing. The change in the bias current changes the Vgs of the feedback amplifier and adjusts the input voltage. The voltage stabilization can be seen in the following diagram.

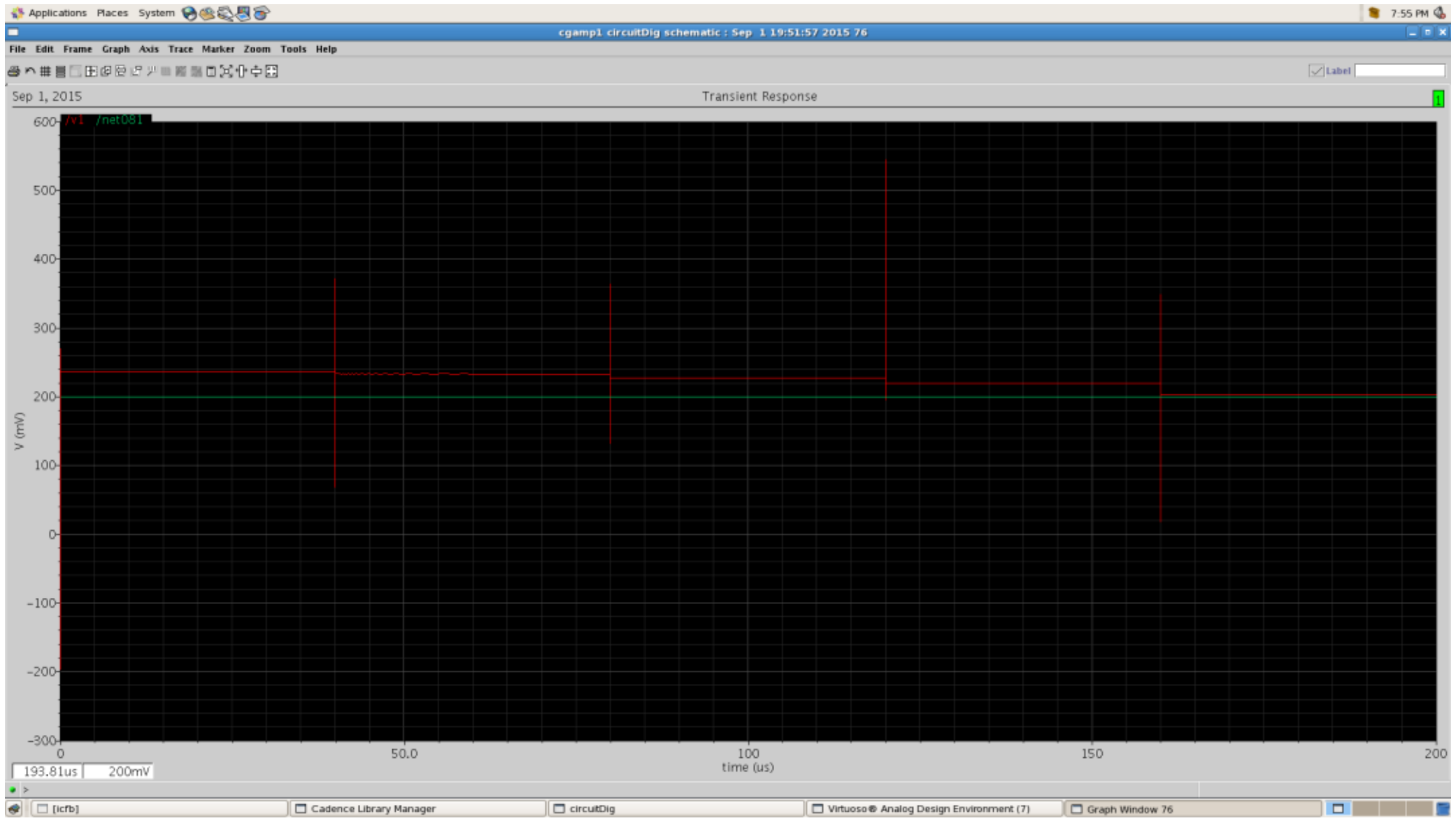

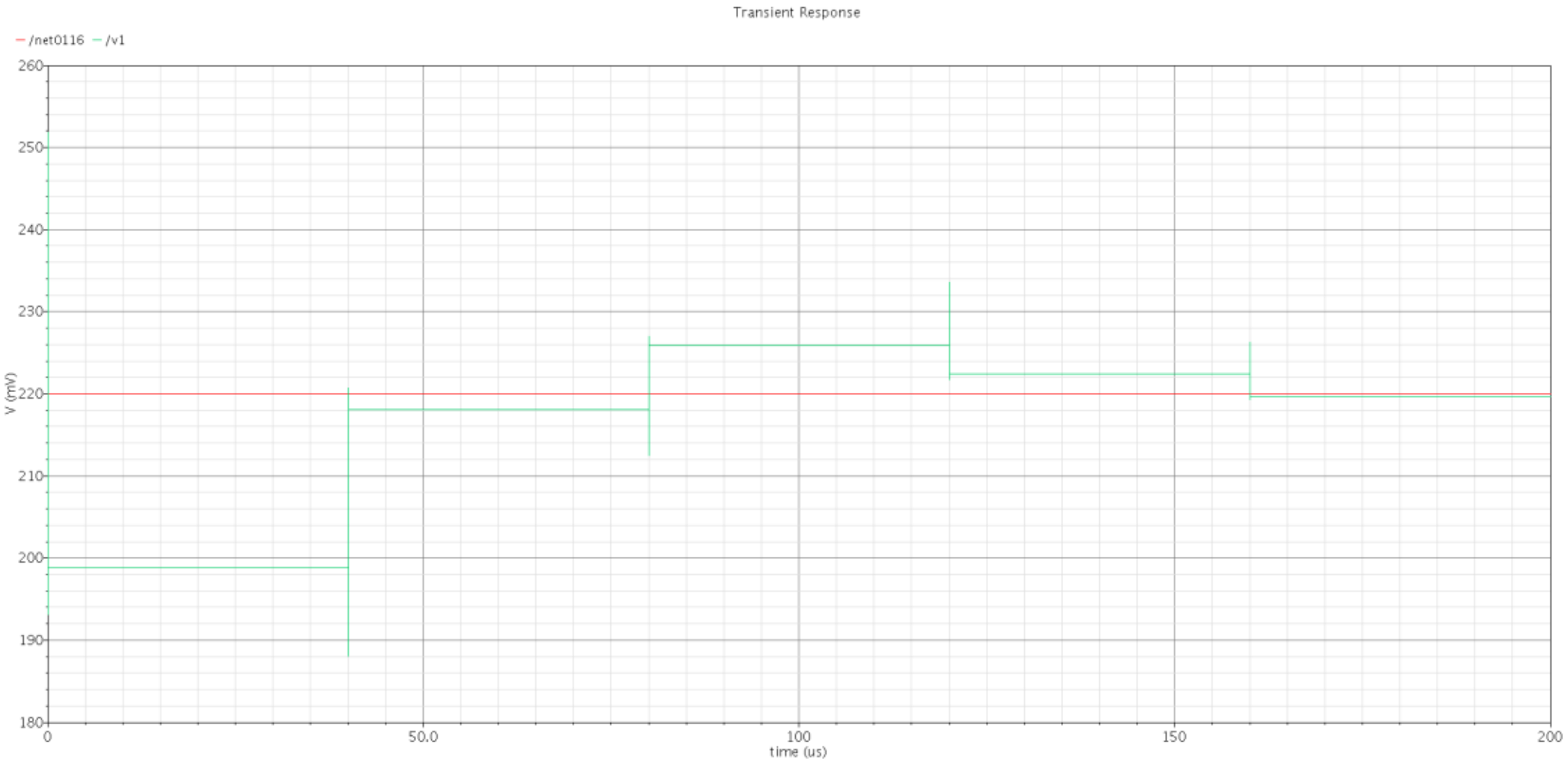

The variability in the DC point due to process variations and mismatch in the transistors was simulated using Monte Carlo analysis in Cadence. The results were checked for the circuit with and without SAR stabilization. As shown in the following figures, the standard deviation in the DC point showed a remarkable drop to 2.63019mV and 2.62507mV from the initial 10.3519mV. The number of runs was 100 and 500 respectively. Thus the variability is considerably reduced and the stable DC point allows better comparison.

Power consumption in this circuit is measured to be 43uW with a Vdd of 1V. The bias current through the main transistor is 5uA.

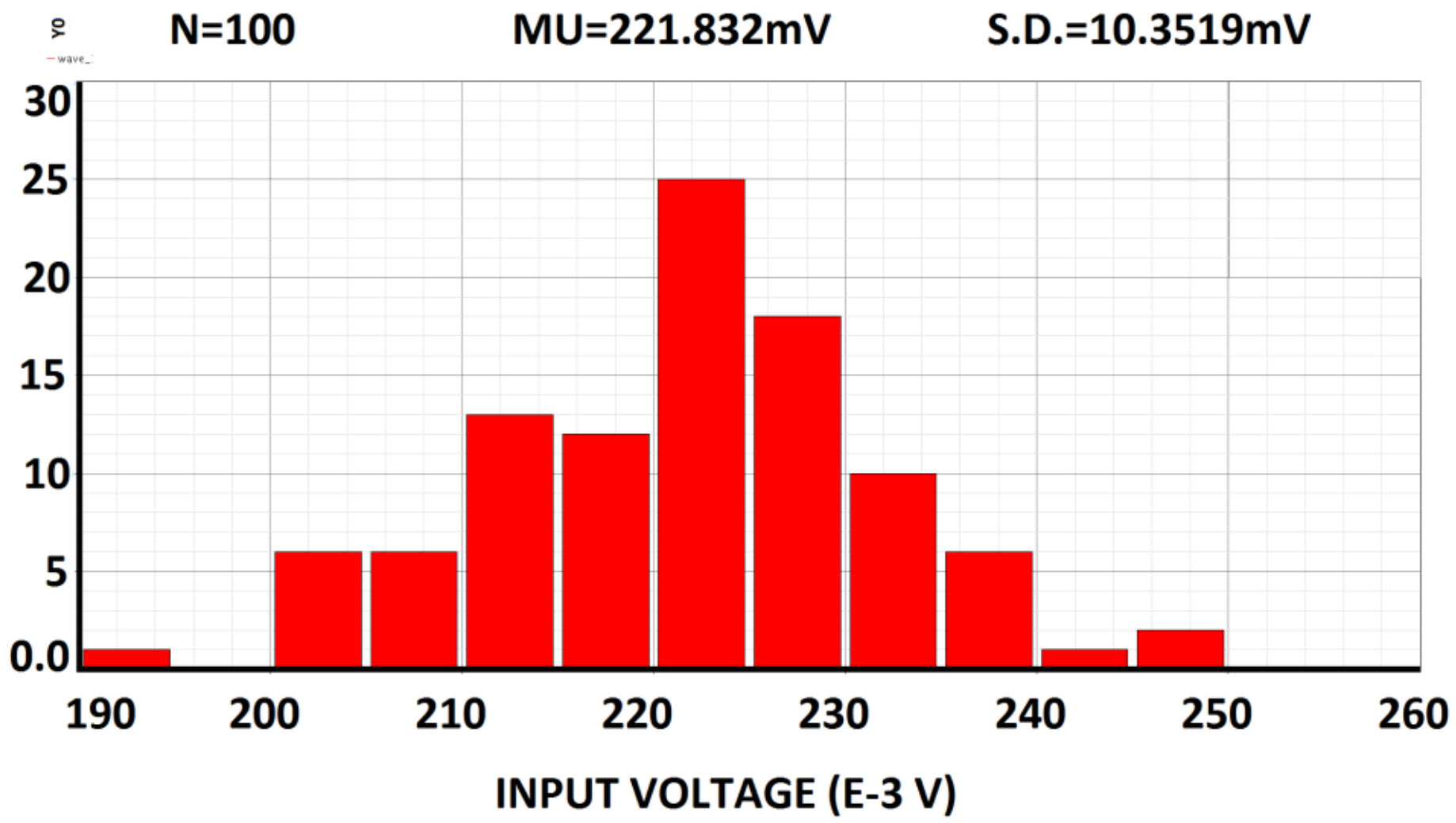

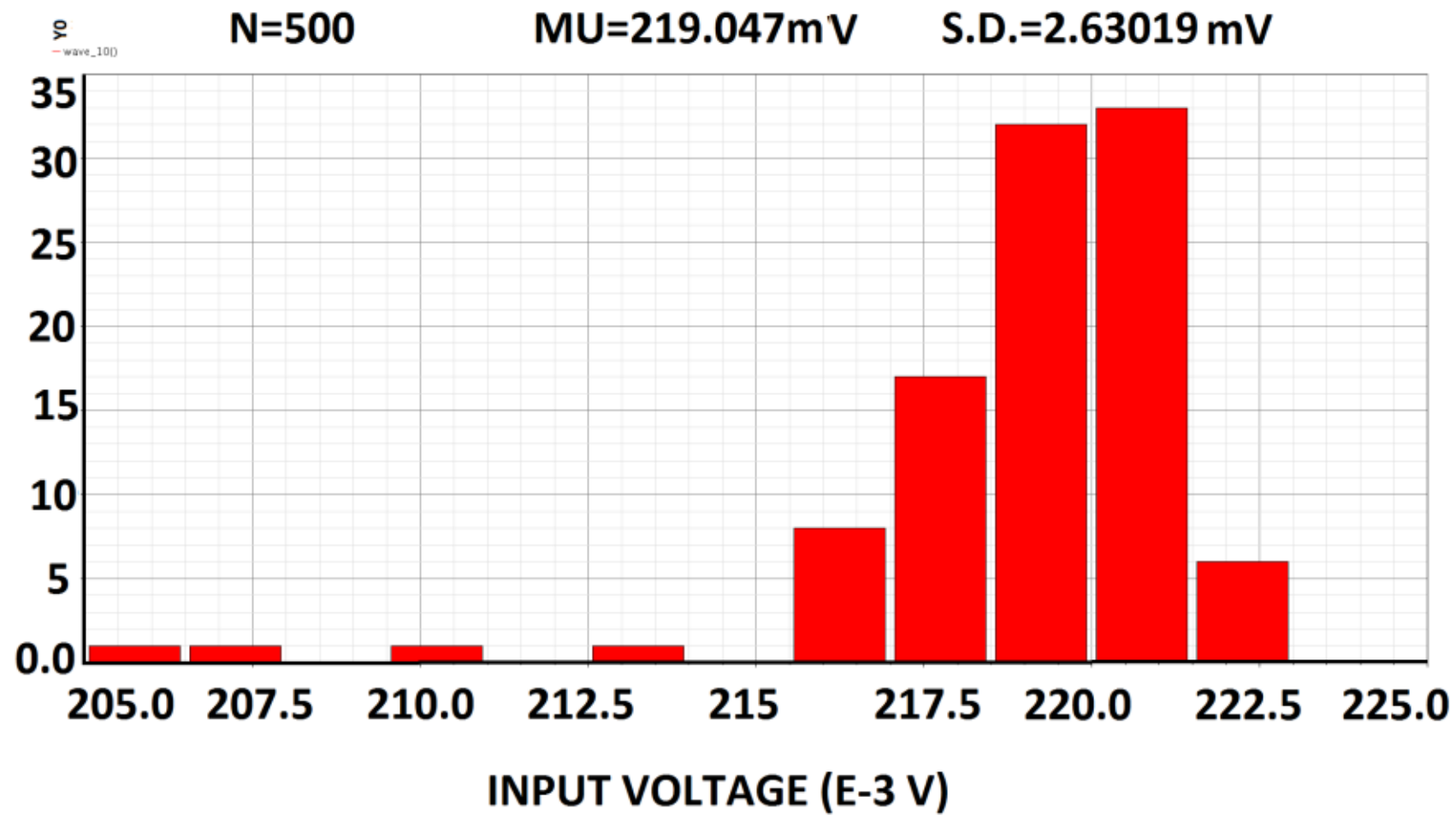

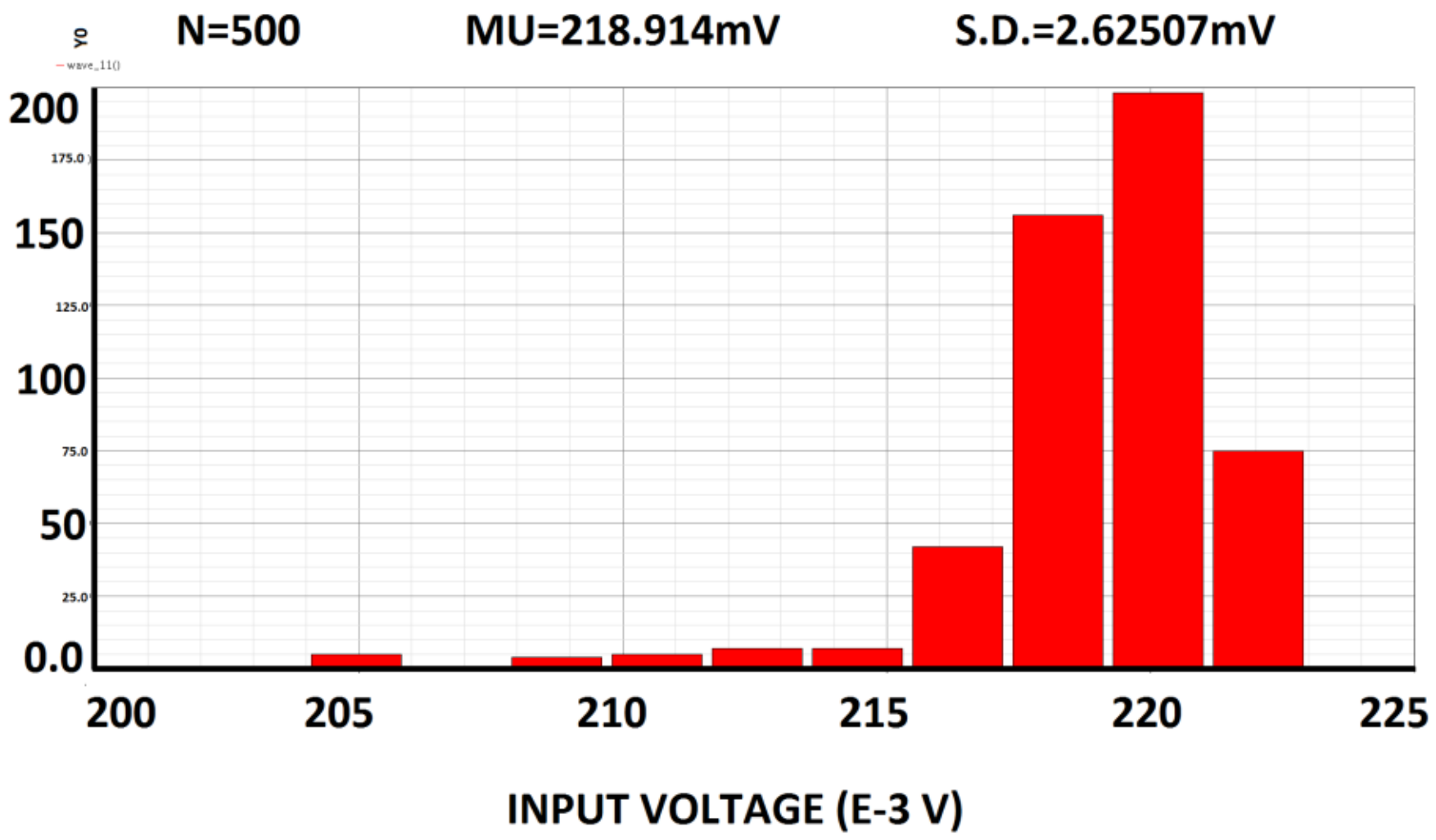

The input and output characteristics are shown below: The response is that of a linear neuron circuit. The bandwidth is also very high, since the circuit effortlessly shows nanosecond response time.

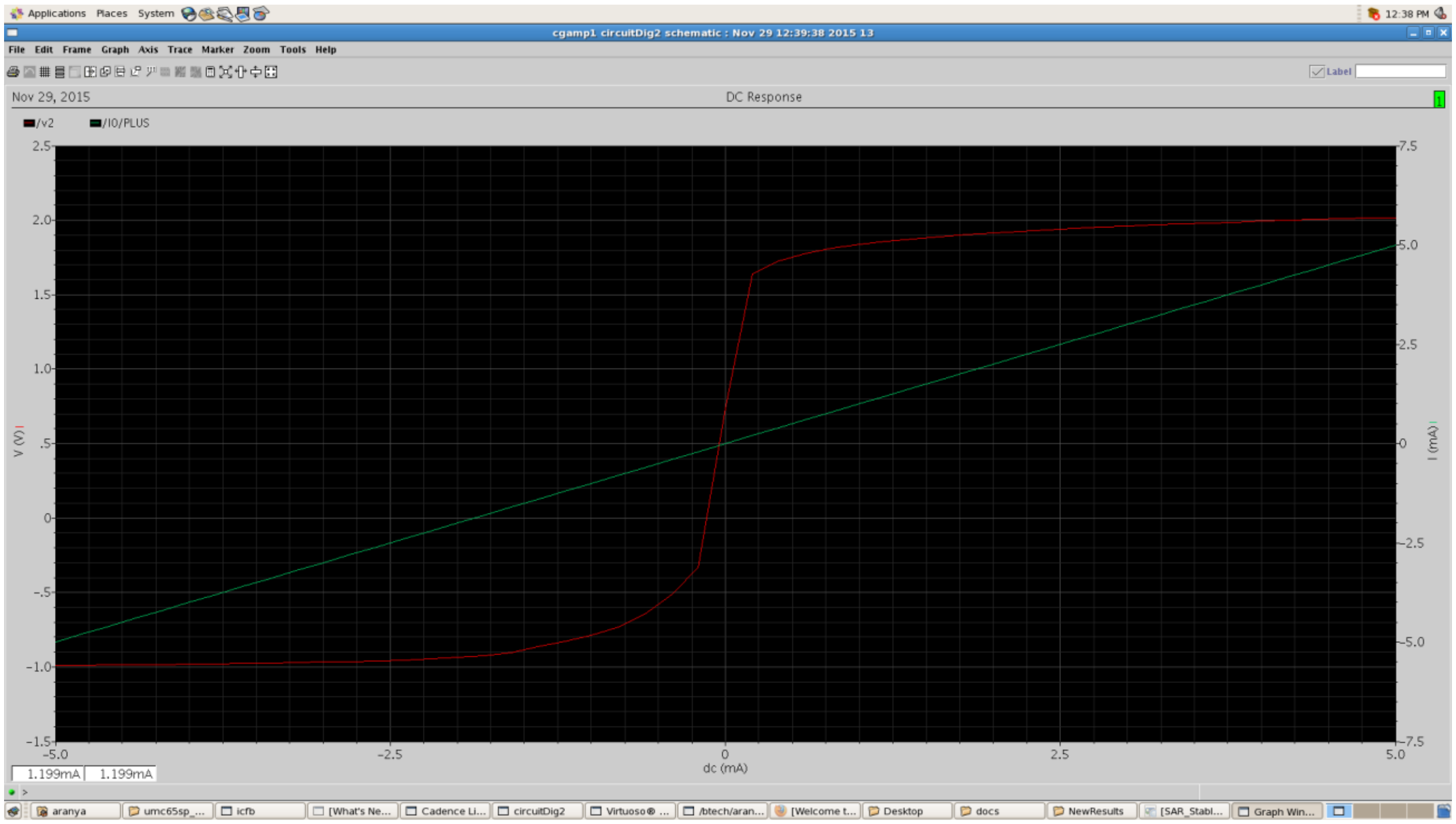

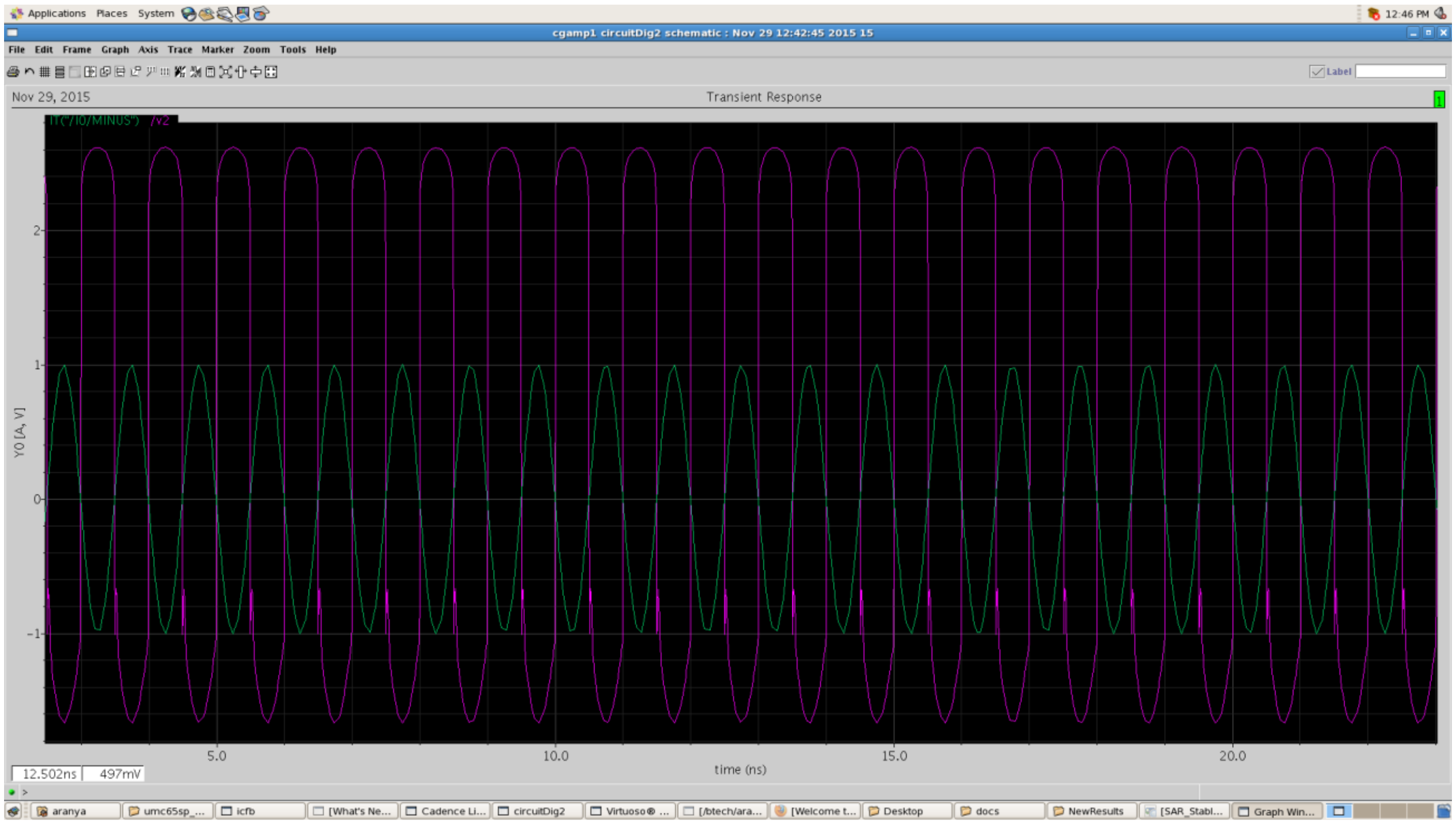

The neuron model with SAR stabilized input and the complete circuit showing two neurons and a shared SAR logic at the input and output is shown below (Cadence Schematic).

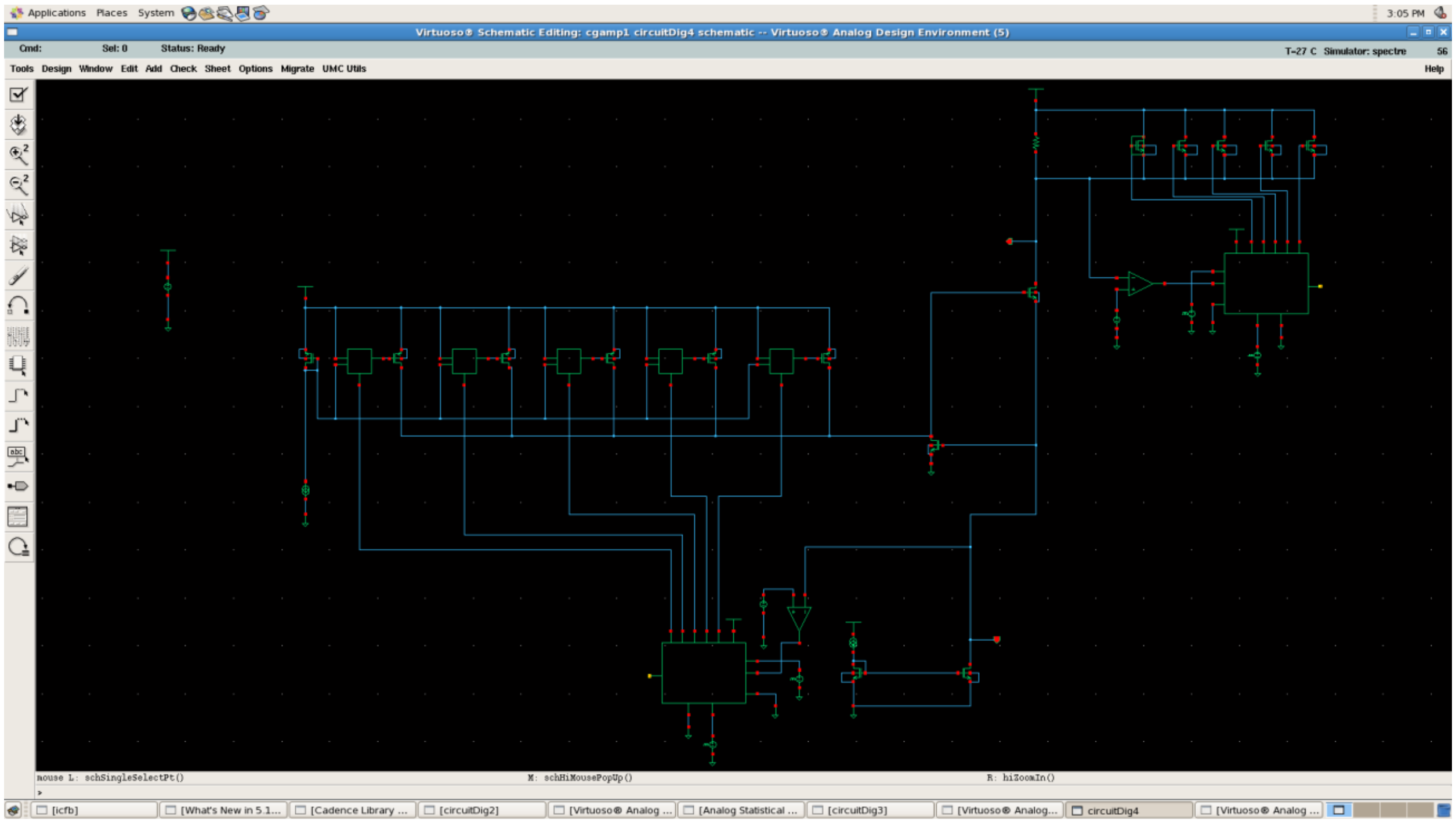

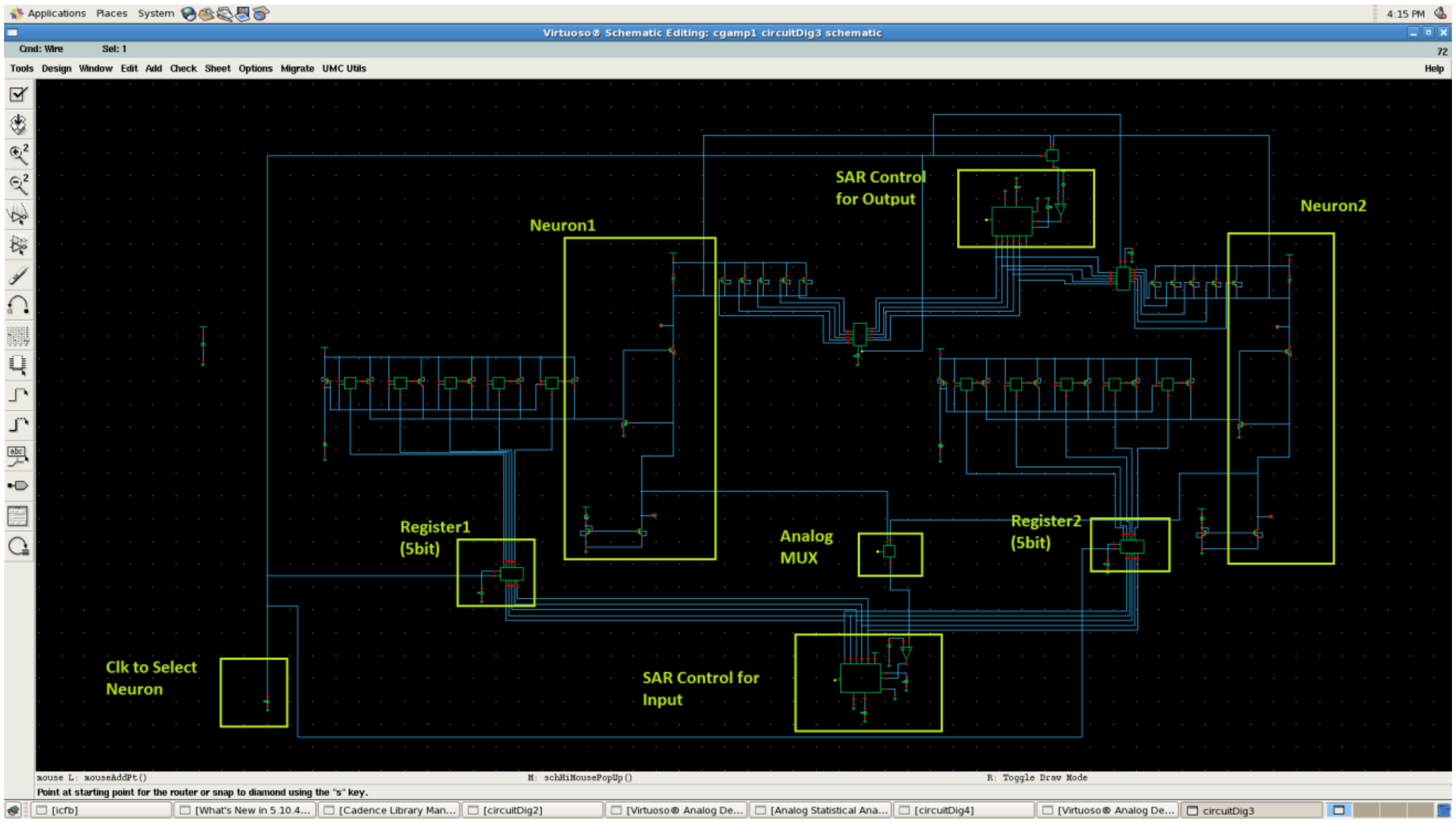

## 3. Conclusion and Future Work

The designed circuit was successful in achieving a low power, low input impedance, DC point stabilization and high bandwidth, characteristics which are ideal for it to be used as the interface between the crossbar structure and the output.

The next level of this project involves preparation of layout diagram of the circuit and fabricating the circuit using commercial cleanroom facilities. Thereafter testing and validation of the circuit will be performed.

[add here…]